\documentclass[11pt,letter]{article}
\usepackage{geometry} 
\usepackage[english]{babel}
\usepackage[latin1]{inputenc}
\usepackage[T1]{fontenc}
\usepackage[mathscr]{euscript}
\usepackage{fullpage}
\usepackage{graphicx}
\usepackage{amssymb,amsmath}
\usepackage{epstopdf}

\title{Non-Clairvoyant Batch Sets Scheduling: \\
Fairness is Fair enough\\ {\Large (Regular Submission) }}
\author{Julien Robert%
\thanks{CNRS -- École Normale Supérieure de Lyon, 46 allée d'Italie, 69364 Lyon Cedex 07, France. \{julien.robert, nicolas.schabanel\}@ens-lyon.fr}$^{~,}$%
\thanks{CNRS -- Centro de Modelamiento Matem\'atico, Blanco Encalada 2120 Piso 7, Santiago de Chile.}\and Nicolas Schabanel$^{*,\dagger}$}
\newcommand{\Sc}[1]{\ensuremath{\mathscr{#1}}}

\newcommand{\algonom}[1]{{\textsc{\bfseries{#1}}}\/}
\newcommand{\EQ}{{\algonom{Equi}}}
\newcommand{\EE}{{\EQ\ensuremath{\circ}\EQ}}
\newcommand{\EA}{{\EQ\ensuremath{\circ}$A$}}
\newcommand{\OF}[1]{\ensuremath{\operatorname{\mathsf{#1}}}}
\newcommand{\MS}{\OF{Makespan}}
\newcommand{\FT}{\OF{Flowtime}}
\newcommand{\SFT}{\OF{Setflowtime}}
\newcommand{\ie}{\textit{i.e.}\/}
\newcommand{\eg}{\textit{e.g.}\/}

\newcommand{\OPT}{{\ensuremath{\operatorname{OPT}}}}

\newcommand{\sch}{{\Sc{S}}}
\newcommand{\schE}{{\Sc{E}}}
\newcommand{\schO}{{\Sc{O}}}

\newcommand{\classOfinst}[1]{{\ensuremath{\mathsf{#1}}}}
\newcommand{\SPS}{\classOfinst{(Par\text{-}Seq)^*}}
\newcommand{\PSP}{\SPS}
\newcommand{\PS}{\classOfinst{Par\text{-}Seq}}
\newcommand{\SP}{\classOfinst{Seq\text{-}Par}}
\newcommand{\Par}{\ensuremath{\operatorname{\mathsf{par}}}}
\newcommand{\Seq}{\ensuremath{\operatorname{\mathsf{seq}}}}

\renewcommand{\phi}{\varphi}

\renewcommand{\leq}{\leqslant}
\renewcommand{\geq}{\geqslant}

\usepackage{theorem}
\newtheorem{theorem}{Theorem}

\newtheorem{lemma}[theorem]{Lemma}
\newtheorem{proposition}[theorem]{Proposition}
\newtheorem{fact}[theorem]{Fact}

\theorembodyfont{\upshape}
\newtheorem{example}{Example}
\newenvironment{proof}{\paragraph{Proof.}}{\hfill$\Box$}
\begin{document}

\maketitle

\begin{abstract}
Scheduling has been since the very beginning a central issue in computer science.
Scheduling questions arise naturally in many different areas among which
operating system design, compiling, memory management, communication network,
parallel machines, clusters management,...  In
real life systems, the characteristics of the jobs (such as release time,
processing time,...) are usually unknown and/or unpredictable beforehand. In particular, the system is
typically unaware of the remaining work in each job or of the ability of the job
to take advantage of more resources. Following these observations, we adopt the job model by Edmonds \emph{et al} (2000, 2003) in which the jobs go through a sequence of different phases. Each phase consists of a certain quantity of work with a different speed-up function that models how it takes advantage of the  number of processors it receives. 
In this paper, we consider non-clairvoyant online setting where a collection of jobs arrives at time $0$. Non-clairvoyant means that the algorithm is unaware of the phases each job goes through and is only aware that a job completes at the time of its completion. 
We consider the
metrics \emph{setflowtime} that was introduced by Robert \emph{et al} (2007). The goal is to minimize
the sum of the completion time of the sets, where a set is completed when all of
its jobs are done. If the input consists of a single set of jobs, the
setflowtime is simply the makespan of the jobs; and if the input consists of a
collection of singleton sets, the setflowtime is simply the flowtime of the
jobs. 
The setflowtime covers thus a continuous range of objective functions from
makespan to flowtime. 
We show that the non-clairvoyant strategy \EE\/ that
evenly splits the available processors among the still unserved sets and then
evenly splits these processors among the still uncompleted jobs of each unserved
set, achieves a competitive ratio $(2+\sqrt3+o(1))\frac{\ln n}{\ln\ln n}$ for
the setflowtime minimization and that this competitive ratio is asymptotically optimal (up to a
constant factor), where $n$ is the size of the largest set. In the special case of a single set, we show that the
non-clairvoyant strategy \EQ\/ achieves a competitive ratio of
$(1+o(1))\frac{\ln n}{\ln\ln n}$ for the makespan minimization problem, which is
again asymptotically optimal (up to a constant factor).
This result shows in particular that as opposed to what previous studies on malleable jobs may let believe, the assertion ``\EQ\/ never starves a job'' is at the same time true and false: false, because we show that it can delay some jobs up to a factor $\frac{\ln n}{\ln\ln n}$, and true, because we show that no algorithm (deterministic or randomized) can achieve a better stretch than $\frac{\ln n}{4\ln\ln n}$.

\end{abstract}

\thispagestyle{empty}

\newpage
\setcounter{page}{1}

\section{Introduction}

Scheduling has been since the very beginning a central issue in computer science. Scheduling questions arise naturally in many different areas among which operating system design, compiling, memory management, communication network, parallel machines, clusters management,... Main contributions to the field go back as far as to the 1950's (\eg, \cite{Smith1956}). It is usually assumed that all the characteristics of the jobs are known at time $0$. It turns out that in real life systems, the characteristics of the jobs (such as release time, processing time,...) are usually unknown and/or unpredictable beforehand. In particular, the system is typically unaware of the remaining work in each job or of the ability of the job to take advantage of more resources. A first step towards a more realistic model was to design algorithms that are unaware of the existence of a given job before its release time \cite{SleatorTarjan1985,KarlinManasseRudolpohSleator1988}. This gave rise to the field of online algorithms. The cost of the solution computed by an online algorithm is measured with respect to an optimal solution which is aware of the release dates; the maximum value of the ratio of these two costs is called the \emph{competitive ratio} of the algorithm. Later on, \cite{MotwaniPhilippsTorng1994} introduced the concept of non-clairvoyant algorithm in the sense that the algorithm is unaware of the processing time of the jobs at the time they are released. They show that for flowtime minimization, the competitive ratio of any non-clairvoyant deterministic algorithm is at least $\Omega(n^{1/3})$ and that a randomized non-clairvoyant algorithm achieves a competitive ratio of $\Omega(\log n)$. Remarking that lower bounds on competitive ratio relied on overloading the system, \cite{PhillipsSteinTorngWein2002} proposes to compare the algorithm to an optimum solution with restricted resources. This analysis technique, known as resource augmentation, allows \cite{KalyanasundaramPruhs2000} to show that given $(1+\epsilon)$ more processing power, a simple deterministic algorithm achieves a constant competitive ratio. Concerning makespan minimization in this setting, earlier work by \cite{Graham1966,Graham1969} already conformed to these restrictions and show that the competitive ratio of non-clairvoyant list scheduling is essentially $2$ which is optimal; \cite{FeldmannKaoSgallTeng1998} proposes as well an optimal algorithm when there exists precedence constraints, with competitive ratio $2.6180$.
Extensive experimental studies (e.g., \cite{LeuteneggerVernon1990,ChiangMansharamaniVernon1994}) have been conducted on various scheduling heuristics. It turns out that real jobs are not fully parallelizable and thus the models above are not adequate in practice. To refine the model,  \cite{DBLP:journals/scheduling/EdmondsCBD03,Edmonds1999} introduce a very general setting for non-clairvoyance in which the jobs go through a sequence of different phases. Each phase consists of a certain quantity of work with a speed-up function  that models how it takes advantage of the  number of processors it receives. For example, during a fully parallel phase, the speed-up function increases linearly with the number of processors received. They prove that even if the scheduler is unaware of the characteristics of each phase, some policies achieve constant factor approximation of the optimal flowtime. More precisely, in \cite{DBLP:journals/scheduling/EdmondsCBD03}, the authors show that the \EQ\/ policy, introduced in the 1980's by \cite{TuckerGupta1989} and implemented in a lot of real systems, achieves a competitive ratio of $(2+\sqrt3)$ for flowtime minimization when all the jobs arrive at time $0$. \cite{Edmonds1999} shows that in this setting
no non-clairvoyant scheduler can achieve a competitive ratio better than $\Omega(\sqrt n)$ when jobs arrive at arbitrary time and shows that \EQ\/ achieves a constant factor approximation of the optimal flowtime if it receives slightly more than twice as much resources as the optimal clairvoyant schedule it is compared to. We refer the reader to the survey \cite{Handbook2007} for a current state of the field. It turns out that in real life systems, the characteristics of the jobs (such as release time, processing time,...) are usually unknown and/or unpredictable beforehand. In particular, the system is typically unaware of the remaining work in each job or of the ability of the job to take advantage of more resources. 

In this paper, we adopt the job model of \cite{DBLP:journals/scheduling/EdmondsCBD03,Edmonds1999} and consider the metrics \emph{setflowtime} that was introduced by \cite{RobertSchabanel2007} in the context of data broadcast scheduling with dependencies. We consider the case where a collection of sets of jobs arrive at time $0$. The goal is to minimize the sum of the completion time of the sets, where a set is completed when all of its jobs are done. If the input consists of a single set of jobs, the setflowtime is simply the makespan of the jobs; and if the input consists of a collection of singleton sets, the setflowtime is simply the flowtime of the jobs. The setflowtime covers thus a continuous range of objective functions from makespan to flowtime. This metrics introduces a minimal form of dependencies between jobs of a given set. In the special case where jobs consist of a single sequential phase followed by a fully parallel phase (with arbitrary release dates), \cite{RobertSchabanel2007} shows that the competitive ratio of the non-clairvoyant strategy \EA\/, that splits evenly the processors among the uncompleted set of jobs and schedules the uncompleted jobs of the set within these processors according to some algorithm $A$, is $O(1)$ with constant resource augmentation.

As in \cite{DBLP:journals/scheduling/EdmondsCBD03}, we focus in this article on the case where all the sets of jobs are released at time $0$, a typical situation of a high performance cluster that receives all the jobs from different members of an institution at the time the institution is granted the access to the cluster. We show that the non-clairvoyant strategy \EE\/ that evenly splits the available processors among the still unserved clients and then evenly splits these processors among the still uncompleted jobs of each unserved client, achieves a competitive ratio $(2+\sqrt3+o(1))\frac{\ln n}{\ln\ln n}$ for the setflowtime minimization and that it is asymptotically optimal (up to a constant factor), where $n$ is the size of the largest set (Theorem~\ref{thm:EE}). In the special case of a single set, we show that the non-clairvoyant strategy \EQ\/ achieves a competitive ratio of $(1+o(1))\frac{\ln n}{\ln\ln n}$ for the makespan minimization problem, which is again asymptotically optimal (up to a constant factor) (Theorem~\ref{thm:EQ}). This result shows that as opposed to what previous studies on malleable jobs may let believe, the assertion ``\EQ\/ never starves a job'' is at the same time true and false: false, because we show that it can delay some jobs up to a factor $\frac{\ln n}{\ln\ln n}$, and true, because we show that no algorithm (deterministic or randomized) can achieve a better stretch than $\frac{\ln n}{4\ln\ln n}$.

As a byproduct of our analysis, we extend the reduction shown by Edmonds in \cite[Lemma~1]{Edmonds1999}.  We show that in order to analyze the competitiveness of a non-clairvoyant scheduler in the general job phase model, one only needs to consider jobs consisting of sequential or parallel work \emph{whatever} the objective function is (flowtime, makespan, setflowtime, stretch, energy consumption,...) (Proposition~\ref{prop:red:SPS}). This last result demonstrates that these two regimes are of the highest interest for the analysis of non-clairvoyant schedulers since they are much easier to handle and allows to treat the very wide range of non-decreasing sublinear speed-up functions all at once.

\smallskip

The next section introduces the model and the notations. Section~\ref{sec:red:SPS} extends the reduction to jobs with sequential or parallel phases, originally proved by \cite{Edmonds1999}. Section~\ref{sec:EQ} shows that \EQ\/ achieves an asymptotically optimal competitive ratio for non-clairvoyant makespan minimization, and introduces the tools that will be used in the last section to obtain the competitiveness of \EE\/ for non-clairvoyant setflowtime minimization.

\section{Non-clairvoyant Batch Sets Scheduling}

\paragraph{The problem.} 
We consider a collection $S=\{S_1,\ldots,S_m\}$ of sets $S_i=\{J_{i,1},\ldots,J_{i,n_i}\}$ of $n_i$ jobs, each of them arriving at time zero. 
A \emph{schedule} $\sch_p$ on $p$ processors is a set of piecewise constant functions%
\footnote{Requiring the functions $(\rho_{ij})$ to be piecewise constant is not restrictive since any finite set of reasonable (\ie, Riemann integrable) functions can be uniformly approximated from below within an arbitrary precision by piecewise constant functions. In particular, all of our results hold if $\rho_{ij}$ are  piecewise continuous functions.}  
$\rho_{ij}:t\mapsto \rho_{ij}^t$ where $\rho_{ij}^t$ is the \emph{amount of processors} allotted to job $J_{ij}$ at time $t$; $(\rho_{ij}^t)$ are arbitrary non-negative real numbers, such that at any time: $\sum_{i,j} \rho_{ij}^t \leq p$. 
Following the definition introduced by~\cite{DBLP:journals/scheduling/EdmondsCBD03}, each job $J_{ij}$ goes through a series of \emph{phases} $J_{ij}^1,\ldots, J_{ij}^{q_{ij}}$ with different degree of parallelism; the amount of \emph{work} in each phase $J_{ij}^k$ is $w_{ij}^k$; at time~t, during its $k$-th phase, job $J_{ij}$ progresses at a \emph{rate} given by a \emph{speed-up function} $\Gamma_{ij}^k(\rho_{ij}^t)$ of the amount $\rho_{ij}^t$ of processors allotted to $J_{ij}$, that is to say that the amount of work accomplished between $t$ and $t+dt$ during phase $J_{ij}^k$ is $\Gamma_{ij}^k(\rho_{ij}^t) dt$. Let $t_{ij}^k$ denote the completion time of the $k$-th phase of $J_{ij}$, \ie\/ $t_{ij}^k$ is the first time $t'$ such that $\int_{t_{ij}^{k-1}}^{t'} \Gamma_{ij}^k(\rho_{ij}^t)\,dt = w_{ij}^k$ (with $t_{ij}^0=0$). Job $J_{ij}$ is completed at time $c_{ij}= t_{ij}^{q_{ij}}$. A schedule is \emph{valid} if all jobs eventually complete, \ie, $c_{ij}<\infty$ for all $i,j$. Set $S_i$ is completed at time $c_i = \max_{j=1..n_i} c_{ij}$. 
The \emph{flowtime} of the jobs in a schedule $\sch_p$ is: $\FT(\sch_p) = \sum_{i,j} c_{ij}$. The \emph{makespan} of the jobs in $\sch_p$ is: $\MS(\sch_p) = \max_{i,j} c_{ij}$. The \emph{setflowtime} of the sets in $\sch_p$ is: $\SFT(\sch_p)=\sum_{i=1}^m c_{i}$. Note that: if the input collection $S$ consists of a single set $S_1$, the setflowtime of a schedule $\sch_p$ is simply the makespan for the jobs in $S_1$; and if $S$ is a collection of singleton sets $S_i=\{J_{i\,1}\}$, the setflowtime of $\sch_p$ is simply the flowtime of the jobs. The setflowtime allows then to measure a continuous range of objective functions from makespan to flowtime. Our goal is to minimize the setflowtime of a collection of sets of jobs arriving at time $0$.

We denote by $\OPT_p(S)$ (or simply $\OPT_p$ or $\OPT$ if the context is clear) the optimal setflowtime of a valid schedule on $p$ processors for collection $S$: $\OPT_p = \inf_{\text{all schedules $\sch_p$}} \SFT(\sch_p)$.

\paragraph{Speed-up functions.} We make the following reasonable assumptions on the speed-up functions. In the following, we consider that each speed-up function is \emph{non-decreasing} and \emph{sub-linear} (\ie, such that for all $i,j,k$, $\rho<\rho' \Rightarrow \frac{\Gamma_{ij}^k(\rho)}{\rho}\geq \frac{\Gamma_{ij}^k(\rho')}{\rho'}$). These assumptions are usually verified (at least desirable...) in practice: non-decreasing means that giving more processors cannot deteriorate the performances; sub-linear means that a job make a better use of fewer processors: this is typically true when parallelism does not take too much advantage of local caches. As shown in \cite{Edmonds1999}, two types of speed-up functions will be of particular interest here: the \emph{sequential} phase where $\Gamma(\rho)= 1$ for all $\rho \geq 0$ (the job progresses at constant speed even if no processor is allotted to it, similarly to an idle period); and the \emph{fully parallel} phase where $\Gamma(\rho) = \rho$ for all $\rho\geq 0$.
Two classes of instances will be useful in the following. We denote by \SPS\/ the class of all instances in which each phase of each job is either sequential or fully parallel, and by \PS\/ the class of all instances in which each job consists of a fully parallel phase followed by a sequential phase. Given a \SPS\/ job $J$, we denote by $\Par(J)$ (resp., $\Seq(J)$) the sum of the fully parallel (resp., sequential) works over all the phases of $J$. Given a set $S_i=\{J_{i,1},\ldots,J_{i,n_i}\}$  of \SPS\/ jobs, we denote by $\Par(S_i) = \sum_{j=1}^n \Par(J_{ij})$ and $\Seq(S_i) =  \max_{j=1,\ldots,n_i} \Seq(J_{ij})$.

\paragraph{Non-clairvoyant scheduling.} In a real life system, the scheduler is typically not aware of the speedup functions of the jobs, neither of the amount of work that remains for each job. Following the definition in~\cite{DBLP:journals/scheduling/EdmondsCBD03,Edmonds1999}, we consider the \emph{non-clairvoyant} setting of the problem. In this setting, the scheduler knows nothing about the progress of each job and is only informed that a job is completed \emph{at the time of its completion}. In particular, it is not aware of the different phases that the job goes through (neither of the amount of work nor of the speed-up function). It follows that even if all the job sets arrive at time $0$, the scheduler has to design an \emph{online strategy} to adapt its allocation on-the-fly to the overall progress of the jobs. We say that a given scheduler $A_p$ is \emph{$c$-competitive} if it computes a schedule $A_p(S)$ whose setflowtime is at most $c$ times the optimal \emph{clairvoyant} setflowtime (that is aware of the characteristics of the  phases of each job), \ie, such that $\SFT(A_p(S))\leq c\cdot\OPT_p(S)$ for all instances $S$. Due to the overwhelming advantage granted to the optimum which knows all the hidden characteristics of the jobs, it is sometimes necessary for obtaining relevant informations on an non-clairvoyant algorithm to limit the power of the optimum by reducing its resources. We say that a scheduler $A_p$ is \emph{$s$-speed $c$-competitive} if it computes a schedule $A_{sp}(S)$ on $sp$ processors whose setflowtime is at most $c$ times the optimal setflowtime on $p$ processors only, \ie, such that $\SFT(A_{sp}(S)) \leq c\cdot \OPT_p(S)$ for all instances~$S$.    

\medskip

We analyse two non-clairvoyant schedulers, namely $\EQ$ and $\EE$, and show that they have an optimal competitive ratio up to constant multiplicative factors. The following two theorems are our main results and are proved in Propositions~\ref{prop:EQ}, \ref{prop:comp:ratio} and \ref{prop:EE}.

\begin{theorem}[Makespan minimization]
\label{thm:EQ}
\EQ\/ is a $\frac{(1+o(1))\ln n}{\ln\ln n}$-competitive non-clairvoyant algorithm for the makespan minimization of a set of $n$ jobs arriving at time $t=0$. Furthermore, no non-clairvoyant deterministic (resp. randomized) algorithm is $s$-speed $c$-competitive for any $s=o(\frac{\ln n}{\ln\ln n})$ and $c < \frac{\ln n}{2\ln\ln n}$ (resp. $c < \frac{\ln n}{4\ln\ln n}$). %
\end{theorem}

\begin{theorem}[Main result]
\label{thm:EE}
\EE\/ is a $\frac{(2+\sqrt3+o(1))\ln n}{\ln\ln n}$-competitive non-clairvoyant algorithm for the setflowtime minimization of a collection of sets of jobs arriving at time $t=0$, where $n$ is the maximum cardinality of the sets. (Clearly the lower bound on competitive ratio given above holds as well for this problem).
\end{theorem}

\section{Non-clairvoyant scheduling reduces to \SPS\/ instances}
\label{sec:red:SPS}

In \cite{Edmonds1999}, Edmonds shows that for the flowtime objective function, one can reduce the analysis of the competitiveness of non-clairvoyants algorithm to the instances composed of a sequence  of infinitesimal sequential or parallel work. It turns out that as shown in Proposition~\ref{prop:red:SPS} below, his reduction is far more general and applies to \emph{any} reasonable objective function (including makespan, setflowtime, stretch, energy consumption,...), and furthermore reduces the analysis to instances where jobs are composed of a finite sequence of positive sequential or fully parallel work, \ie, to \emph{true} \SPS\/ instances.

Consider a collection%
\footnote{Note that the reduction to \SPS\/ instances  applies as well to jobs with release dates, precedences constraints, or any other type of constraints, since Lemma~\ref{lem:red:SPS} simply consists in remapping the phases of the jobs within two \emph{valid} schedules that naturally satisfy these additional constraints.}
of $n$ jobs $J_1,\ldots, J_n$ where $J_i$ consists of a sequence of phases $J_i^1,\ldots,J_i^{q_i}$ of work $w_i^1,\ldots,w_i^{q_i}$ with speed-up functions $\Gamma_i^1,\ldots,\Gamma_i^{q_i}$. Consider a speed $s>0$. Let $A_{sp}$ be a arbitrary non-clairvoyant scheduler on $sp$ processors, and $\schO_p$ a valid schedule of $J_1,\ldots,J_n$ on $p$ processors.

\begin{lemma}[Reduction to \SPS\/ instances] \label{lem:red:SPS} 
There exists a collection of \SPS\/ jobs $J'_1,\ldots,J'_n$ such that $\schO_p[J'/J]$  is a valid schedule of $J'_1,\ldots,J'_n$ and ${A_{sp}(J') = A_{sp}(J)[J'/J]}$, where $\sch[J'/J]$ denotes the schedule obtained by  scheduling job $J'_i$ instead of $J_i$ in a schedule $\sch$.
\end{lemma}

\begin{proof}
The present proof only simplifies the proof originally given in \cite{Edmonds1999} in the following ways: the jobs $J'_1,\ldots,J'_n$ consist of a \emph{finite} number of phases (and are thus a valid finitely described instance), and the schedules computed by algorithm $A_{sp}$ on instances $J'_1,\ldots,J'_n$ and $J_1,\ldots,J_n$ are identical, which avoids to consider infinitely many schedules to construct $J'$ from $J$.  

Consider the two schedules $A_{sp}(J)$ and $\schO_p$. Consider job $J_1$ (the construction of $J'_i$ is identical for $J_i$, $i\geq 2$). Let $\rho_A(t)$ and $\rho_\schO(t)$ be the number of processors allotted overtime to $J_1$ by $A_{sp}(J)$ and $\schO_p$ respectively.  Let $\phi(t)$ be the time $t'$ at which the portion of work of $J_1$ executed in $\schO_p$ at time $t$, is executed in $A_{sp}(J)$. Let $\Gamma_{t'}$ be the speed-up function of the portion of work of $J_1$ executed in $A_{sp}(J)$ at time $t'$. By construction, for all $t$, the same portion of work $dw$ of $J_1$ is executed between $t$ and $t+dt$ in $\schO_p$ and between $\phi(t)$ and $\phi(t+dt) = \phi(t)+d\phi(t)$ in $A_{sp}(J)$ with the same speed-up function $\Gamma_{\phi(t)}$, thus: $dw = \Gamma_{\phi(t)}(\rho_\schO(t))\, dt = \Gamma_{\phi(t)}(\rho_A(\phi(t)))\,d\phi(t)$; it follows that $\phi$'s derivative is $\phi'(t) = \frac{\Gamma_{\phi(t)}(\rho_\schO(t))}{\Gamma_{\phi(t)}(\rho_A(\phi(t)))}$ ($\geq0$, $\phi$ is an increasing function). $\rho_A(\phi(t))$ and $\rho_\schO(t)$ are (by definition) piecewise constant functions. Let $t_1=0<t_2<\cdots<t_\ell$ such that $\rho_A(\phi(t))$ and $\rho_\schO(t)$ are constant on each time interval $[t_{k},t_{k+1})$ and zero beyond $t_\ell$; let $t'_k = \phi(t_k)$, $\rho_A(t')$ is constant on each time interval $(t'_k,t'_{k+1})$; let $\rho_A^k = \rho_A(t'_k)$ and $\rho_\schO^k = \rho_\schO(t_k)$. By construction, the portion of work of $J_1$ executed by $A_{sp}(J)$ between times $t'_k$ and $t'_{k+1}$, is executed by $\schO_p$ between times $t_k$ and $t_{k+1}$. $J'_1$ consists of a sequence of $(\ell-1)$ phases, sequential or fully parallel depending on the relative amount of processors $\rho_\schO^k$ and $\rho_A^k$  alloted by $\schO_p$ and $A_{sp}(J)$ to $J_1$ during time intervals $[t_k,t_{k+1}]$ and $[t'_k, t'_{k+1}]$ respectively. The $k$-th phase of $J'_1$ is defined as follows:
\begin{itemize}
\item
If $\rho_\schO^k \leq \rho_A^k$, the $k$-th phase of $J'_1$ is a sequential work of $w_k = t'_{k+1}-t'_{k}$.
\item
If $\rho_\schO^k > \rho_A^k$, the $k$-th phase of $J'_1$ is a fully parallel work of $w_k = \rho_A^k \cdot (t'_{k+1}-t'_{k})$.
\end{itemize}
The $k$-th phase of $J'_1$ is designed to fit exactly in the overall amount of processors allotted by $A_{sp}$ to $J_1$ during $[t'_{k},t'_{k+1}]$; thus, since $A_{sp}$ is non-clairvoyant, $A_{sp}(J') = A_{sp}(J)[J'/J]$. Let now verify that the $k$-th phase of $J'_1$ fits in the overall amount of processors allotted by $\schO_p$ to $J_1$ during $[t_{k},t_{k+1}]$. 
\begin{itemize}
\item
If $\rho_\schO^k \leq \rho_A^k$, $w_k = \displaystyle\int_{t'_k}^{t'_{k+1}} \!\! dt' = \int_{t_k}^{t_{k+1}} \!\phi'(t) dt = \int_{t_k}^{t_{k+1}} \frac{\Gamma_{\phi(t)}(\rho_\schO^k)}{\Gamma_{\phi(t)}(\rho_A^k)} dt \leq \int_{t_k}^{t_{k+1}} \!\! dt = t_{k+1}-t_k$ since the $\Gamma_{\phi(t)}$ are non-decreasing functions.
\item
If $\rho_\schO^k > \rho_A^k$, $w_k = \rho_A^k \displaystyle\int_{t'_k}^{t'_{k+1}} \!\! dt' = \rho_A^k \int_{t_k}^{t_{k+1}} \frac{\Gamma_{\phi(t)}(\rho_\schO^k)}{\Gamma_{\phi(t)}(\rho_A^k)} dt \leq \rho_A^k \int_{t_k}^{t_{k+1}} \frac{\rho_\schO^k}{\rho_A^k}dt = \rho_\schO^k \cdot (t_{k+1}-t_k)$, since the $\Gamma_{\phi(t)}$ are sub-linear functions.
\end{itemize}
It follows that in both cases, the $k$-th phase of $J'_1$ can be completed in the space allotted to $J_1$ in $\schO_p$ during $[t_k,t_{k+1}]$.
\end{proof}

\medskip

Consider an arbitrary non-clairvoyant scheduling problem where the goal is to minimize an objective function $F$ over the set of all valid schedules of an instance of jobs $J_1,\ldots,J_n$. Assume that $F$ is \emph{monotonic} in the sense that $F(\sch)\leq F(\sch')$ if $\sch$ and $\sch'$ are two valid schedules of $J_1,\ldots,J_n$ such that for all $i$, $J_i$ receives at any time less processors in $\sch$ than in $\sch'$ (note that since a completed job do not receive processors, this implies that for all $i$, $J_i$ cannot complete in $\sch'$ before it completes in $\sch$). Note that \emph{all} standard objective functions are monotonic: flowtime, makespan, setflowtime, stretch, energy consumption, etc. Then,  

\begin{proposition} \label{prop:red:SPS} 
Any non-clairvoyant algorithm $A^F$ for a monotonic objective function $F$ that is $s$-speed $c$-competitive over \SPS\/ instances, is also $s$-speed $c$-competitive over all instances of jobs going through phases with arbitrary non-decreasing sublinear speed-up functions. 
\end{proposition}

\begin{proof}
Consider a non-\SPS\/ instance $J=\{J_1,\ldots,J_n\}$. Denote by $\OPT^F_p(J)$ the optimal cost for $J$, \ie, $\OPT^F_p(J) = \inf \{F(\sch):\text{$\sch$ is a valid schedule of $J$ on $p$ processors}\}$. Consider an arbitrary small $\epsilon>0$ and $\schO$ a valid schedule of $J$ such that $F(\schO)\leq \OPT^F_p(J)+\epsilon$ (note that we do not need that an optimal schedule exists). Let $J'$ be the \SPS\/ instance given by Lemma~\ref{lem:red:SPS} from $J$, $A^F_{sp}$, and $\schO$. Since $A^F_{sp}(J') = A^F_{sp}(J)[J'/J]$, $F(A^F_{sp}(J)) = F(A^F_{sp}(J'))$. But $A^F_{sp}$ is $s$-speed $c$-competitive for $J'$, so: $F(A^F_{sp}(J)) \leq c\cdot\OPT^F_p(J') \leq c\cdot F(\schO[J'/J]) \leq c\cdot F(\schO) \leq c\,\OPT^F_p(J)+c\,\epsilon$, as $\schO[J'/J]$ is a valid schedule of $J'$  and $F$ is monotonic. Decreasing $\epsilon$ to zero completes the proof.~
\end{proof}

\medskip

It follows that for \emph{any} non-clairvoyant scheduling problem, it is enough to analyse the competitiveness of a non-clairvoyant algorithm on \SPS\/ instances. Sequential and parallel phases are both unrealistic (sequential phases that progress at a constant rate even if they receive no processors are not less legitimate than fully parallel phases which do not exist for real either). Nevertheless, these are much easier to handle in competitive analysis, and Proposition~\ref{prop:red:SPS} guarantees that these two extreme(ly simple) regimes are sufficiently general to cover the range of all possible non-decreasing sublinear functions. We shall from now on consider only \PSP\/ instances.

\section{The single set case}
\label{sec:EQ}

In this section, we focus on the case where the collection $S$ consists of a unique set $S_1=\{J_1,\ldots,J_n\}$. The problem consists thus in minimizing the \emph{makespan} of the set of jobs $S_1$. This problem is interesting on its own and, as far as we know, no competitive non-clairvoyant algorithm was known. Furthermore, the analysis that follows is one of the keys to the main result of the next section.  

\subsection{\EQ\/ Algorithm} 

\EQ\/ is the classic operating system approach to non-clairvoyant scheduling. It consists in giving a equal amount of processors to each uncompleted job (operating systems approximate this strategy by a preemptive round robin policy). Formally, given $p$ processors, if $N(t)$ denotes the number of uncompleted jobs at time $t$, \EQ\/ allots $\rho_i^t = p/N(t)$ processors to each uncompleted job $J_i$ at time~$t$.

In~\cite[Theorem~3.1]{DBLP:journals/scheduling/EdmondsCBD03}, the authors show that \EQ\/ is $(2+\sqrt3)$-competitive for the flowtime of the jobs when all the jobs arrive at time $t=0$. As pointed out in~\cite{Edmonds1999}, the key of the analysis is that the contribution to the flowtime of the sequential phases is independent of the scheduling policy, and thus the performance of the scheduler is measured by its ability to give a sufficiently large amount of processors to the parallel phases. When parallel work is delayed by sequential work with respect to the optimum strategy, the number of uncompleted jobs in a parallel phase increases and \EQ\/ allots more and more processing power to parallel work. It follows that \EQ\/ self-adjusts naturally which yields that it has a constant competitive ratio for flowtime minimization.

\label{sec:eq:ex}
When the objective is to minimize the makespan, the  times at which the sequential phases are scheduled matter because they can be arbitrarily delayed by parallel phases as shown in the following example. 

\begin{example}\label{ex:eq}
Consider $n = \ell^\ell$ jobs arriving at time $0$ on one processor. Between time $t=0$ and $t=1$, a fraction $1-1/\ell$ of the $\ell^\ell$ jobs are in a sequential phase of work $1$ and all of them complete at time $1$; the other $1/\ell$ fraction of the jobs is in a parallel phase of work $1/\ell^\ell$ each; \EQ\/ allots to each job an equal processing power $1/\ell^\ell$ during this time interval and at time $1$ only remains the $\ell^\ell/\ell = \ell^{\ell-1}$ jobs that just finish their first parallel phase.  We continue recursively as follows until time $t=\ell$  as illustrated on Fig.~\ref{fig:bad}: at integer time $t=i<\ell$, $\ell^{\ell-i}$ jobs are still uncompleted; between time $t=i$ and $t=i+1$, a fraction $1-1/\ell$ of the $\ell^{\ell-i}$ jobs are in a sequential phase of work $1$ and all of them complete at time $i+1$; the other $1/\ell$ fraction of the jobs is in a parallel phase of work $1/\ell^{\ell-i}$ each; \EQ\/ allots to each job an equal processing power $1/\ell^{\ell-i}$ during this time interval and at time $i+1$ only remains the $\ell^{\ell-i}/\ell = \ell^{\ell-(i+1)}$ jobs that just finish their $i$-th parallel phase. At time $t=\ell$, there only remains one job which completes at time $\ell+1$ after a sequential phase of work $1$.

\begin{figure}[htb]
\centerline{\includegraphics[height=5cm]{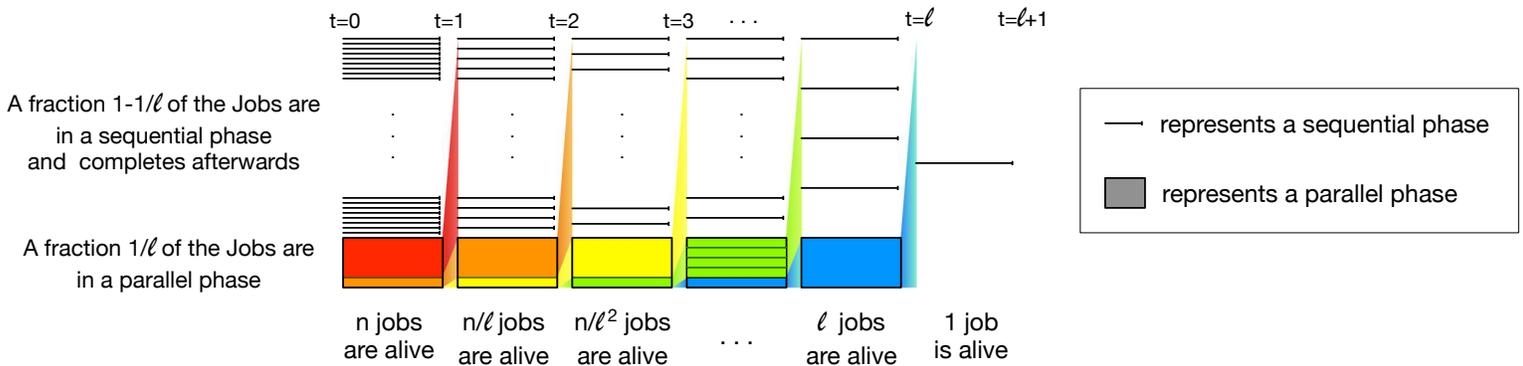}}
\caption{An inefficient execution of \EQ\/. } 
\label{fig:bad}
\end{figure}

It follows that for this instance, \EQ\/ achieves a makespan of $\ell+1$. But, the amount of parallel work executed within each time interval $[i,i+1]$ for $i=0,...,\ell-1$, equals to $1/\ell$. It follows that an optimal (clairvoyant) scheduler can complete all the parallel work in one time unit and then finish the remaining sequential work before time $2$.  Since $n = \ell^\ell$ and $\ell > \frac{\ln n}{\ln\ln n}$, we conclude:
\end{example}

\begin{fact}
\EQ\/ is not $c$-competitive for the makespan minimization problem, for any $c\leq \frac{\ln n}{2\ln\ln n}$.
\end{fact} 

It follows that as opposed to the flowtime minimization, we need to take into account the delay introduced by parallel phases over sequential phases. (Note that for the instance above, the flowtime achieved by \EQ\/ is $\frac{1-1/\ell^{\ell-1}}{1-1/\ell}= 1+1/\ell + o(1/\ell)$ which is asymptotically optimal.)  

\subsection{Analysis of \EQ\/ for makespan minimization}

Thanks to Proposition~\ref{prop:red:SPS}, we focus on a \PSP\/ instance $S=\{J_1,\ldots,J_m\}$. By rescaling the parallel work in each job, we can assume w.l.o.g.\/ that $p=1$. We show that the behavior exhibited in the example of section~\ref{sec:eq:ex} is indeed the worst case behavior of \EQ. Let us define the \PS\/ instance $S'=\{J'_1,...,J'_n\}$ where each $J'_i$ consists of a fully parallel phase of work $\Par(J_i)$ followed by a sequential phase of work $\Seq(J_i)$. Observe that:

\begin{lemma} \label{lem:S<=S'}
$\MS(\EQ(S)) \leq \MS(\EQ(S')).$
\end{lemma}

\begin{proof}
Since all the jobs arrive at time $0$, the number of uncompleted jobs is a non-increasing function of time. It follows that the amount of processors alloted by $\EQ$ to a given job is a non-decreasing function of time. Thus, moving all the parallel work to the front, can only delay the completion of the jobs since less processors will then be allocated to each given piece of parallel work. 
\end{proof}

\begin{proposition} \label{prop:EQ}
\EQ\/ is $(1+o(1))\frac{\ln n}{\ln\ln n}$-competitive for the makespan minimization problem.
\end{proposition}

\begin{proof}
Consider the schedule $\EQ(S')$ and let $T = \MS(\EQ(S'))$. We write $[0,T]$ as the disjoint union of two sets $A$ and $\bar{A}$. Set $\alpha = \frac{(\ln\ln n)^2}{\ln n}$. Recall that  $N(t)$ is the number of uncompleted jobs at time~$t$. Let $s_t$ be the number of uncompleted jobs in a sequential phase at time $t$. Set $A$ is the set of all the instants where the fraction of jobs in a sequential phase is larger than $\alpha$, and $\bar{A}$ is its complementary set: \ie, $A = \{0\leq t\leq T: s_t \geq (1-\alpha) N(t)\}$ and $\bar{A} = \{0\leq t\leq T:s_t < (1-\alpha) N(t)\}$. Clearly, $T = |A| + |\bar{A}|$, with $|X| = \int_X dt$. We now bound $|A|$ and $|\bar A|$ independently.

At any time $t$ in $\bar A$, the total amount of parallel work completed between $t$ and $t+dt$ is at least $\alpha\,dt$. Since the total amount of parallel work is $\Par(S')$, we get $\int_{\bar A} \!\!\alpha\, dt \leq \Par(S')$. Thus, $|\bar A| \leq \Par(S')/\alpha$. 

Now, let $t_1< \cdots < t_q$ with $t_k\in A$ for all $k$, such that the  time intervals $I_1= [t_1,t_1+\Seq(S')),\ldots,I_q = [t_q,t_q+\Seq(S'))$ form a collection of non-overlapping intervals of length $\Seq(S')$ that covers $A$. Once the sequential phase of a \PS\/ job has begun at or before time $t$, the job completes before time $t+\Seq(S')$. Since at time~$t_k$, at least $(1-\alpha) \cdot N(t_k)$  jobs are in a sequential phase, at time $t_{k+1} \geq t_k+\Seq(S')$, we have thus: $N(t_{k+1})\leq \alpha N(t_k)$. It follows that $N(t_k) \leq \alpha^k\cdot n$. Since $N(t_q)\geq 1$, $q\leq \frac{\ln n}{\ln(1/\alpha)}$. But $A$ is covered by $q$ time intervals of length $\Seq(S')$, so: $|A| \leq \frac{\ln n}{\ln(1/\alpha)} \Seq(S')$. Finally, 
\begin{align*}
\MS(\EQ(S)) &\leq \MS(\EQ(S')) = T \\
	&  \leq \textstyle\frac{1}{\alpha} \Par(S') +\frac{\ln n}{\ln (1/\alpha)} \Seq(S')\\
	&\leq \textstyle (1+o(1))\frac{\ln n}{\ln\ln n} \max(\Par(S'),\Seq(S'))\\
	& = \textstyle (1+o(1))\frac{\ln n}{\ln\ln n} \max(\Par(S),\Seq(S))\\
	& \leq \textstyle(1+o(1))\frac{\ln n}{\ln\ln n} \OPT(S).
\end{align*}
\vspace*{-1.5em}\\
\hspace*{\stretch{1}}~
\end{proof}

\subsection{\EQ\/ is asymptotically optimal up to a factor 2}

The following lemma generalizes the example given in section~\ref{sec:eq:ex} and shows that \EQ\/ is asymptotically optimal in the worst case. Note that increasing the number of processors by a factor $s$ does \emph{not}  improve the competitive ratio of any deterministic or randomized algorithm as long as $s = o(\frac{\ln}{\ln\ln n})$, \ie, the competitive ratio does not improve even if the number of processors increases (not too fast)  with the number of jobs.  

\begin{proposition}[Lower bound on the competitive ratio of any non-clairvoyant algorithm] \label{prop:comp:ratio}
No non-clairvoyant algorithm $A$  has a competitive ratio less than $\gamma_D = \frac{\ln n}{2 \ln\ln n}$ if $A$ is deterministic, and $\gamma_R = \frac{\ln n}{4 \ln\ln n}$ if $A$ is randomized.

Furthermore, no non-clairvoyant algorithm $A$ is $s$-speed $c$-competitive for any speed $s=o(\frac{\ln}{\ln\ln n})$ if $c<\gamma_D$ and $A$ is deterministic, or $c<\gamma_R$  if $A$ is randomized.
\end{proposition}

\begin{proof}
We first extend Example~\ref{ex:eq} to cover all deterministic algorithms. Consider the execution of an algorithm $A_s$ given $s$ processors on the following instance. At time $0$, $n = (s \ell)^{\ell}$ jobs are given. Since the algorithm is non-clairvoyant, we set the phase afterwards. At time $1$, we renumber the jobs $J_1,\ldots,J_n$ by non-decreasing processing power received between $t=0$ and $t=1$ in $A_s$. Between time $t=0$ and $t=1$, we set the jobs $J_{(s\ell)^{\ell-1}+1},\ldots,J_n$ (\ie, the last fraction $1-1/(s\ell)$ of the $(s \ell)^{\ell}$ jobs) to be in a sequential phase of work $1$ and say that all of them complete at time $1$; each $J_j$ of the $J_1,\ldots,J_{(s\ell)^{\ell-1}}$ are set in a parallel phase of work $\int_{0}^{1} \rho_j^t\, dt$ each between time $0$ and $1$, where $\rho_j^t$ is the amount of processors alloted to $J_i$ at time $t$. The processing power received by the last $1-1/(s\ell)$ fraction of jobs between $t=0$ and $t=1$ is at least $s-1/\ell$ and thus, the total parallel work assigned to the jobs between $0$ and $1$ is at most $1/\ell$. 
At time $1$ only remains the jobs $J_1,\ldots,J_{(s\ell)^{\ell-1}}$ that just have finished their first parallel phase. 
We continue recursively as follows until time $t=\ell$: at integer time $t=i<\ell$, $(s\ell)^{\ell-i}$ jobs are still uncompleted; between time $t=i$ and $t=i+1$, the fraction $1-1/(s\ell)$ of the $(s\ell)^{\ell-i}$ jobs that received the most processing power are set in a sequential phase of work $1$ and all of them complete at time $i+1$; each job $J_j$ of the other $1/(s\ell)$ fraction is set in a parallel phase of work $\int_{i}^{i+1} \rho_j^t\, dt$  each; At time $i+1$ only remains the $(s\ell)^{\ell-i}/(s\ell) = (s\ell)^{\ell-(i+1)}$ jobs that just have finished their $i$-th parallel phase. At time $t=\ell$, there only remains one job which completes at time $\ell+1$ after a sequential phase of work $1$. It follows that for this instance, $A_s$ achieves a makespan of $\ell+1$. But, the amount of parallel work executed within each time interval $[i,i+1]$ for $i=0,...,\ell-1$, is at most $1/\ell$. It follows that an optimal (clairvoyant) scheduler on $1$ processor can complete all the parallel work in one time unit and then finish the remaining sequential work before time $2$.  But $n = (s\ell)^\ell$,  $\ell > \frac{\ln n}{\ln\ln n}$, which concludes the proof.

We use the Yao's principle (see \cite{Yao1977,MotwaniRaghavan1995}) to extend the result to randomized algorithms. Due to space constraint, we just sketch the proof. Take an arbitrary deterministic scheduler $A$, we will show that $A$ achieves expected makespan of at least $\frac{\ln n}{4\ln\ln n}$ on the random instance obtained by: 1) making $n$ copies of each job in the instance of Example~\ref{ex:eq};  2) dividing the parallel work of each job by $n$; and 3) taking a random permutation of the $n^2$ resulting jobs. Take $\epsilon>0$, at time $1$, at most $\frac{n^2}{1+\epsilon}$ jobs have received at least $\frac{1+\epsilon}{n^2}$. Since $A$ is non-clairvoyant and since the jobs are randomly permuted, the expected number of jobs starting with a parallel phase ($\frac{n^2}{\ell}$ in total) that have received between time $0$ and $1$ at most $\frac{1+\epsilon}{n^2}$ processors is at least $\frac{n^2}{(1+\epsilon)\ell}$. Since the hypergeometric distribution (the distribution given by a permutation, see \cite{Feller}) is more concentrated than the binomial,  the Chernoff bound tells that the complementary probability that at most $\frac{n^2}{2(1+\epsilon)\ell}$ jobs did not complete their parallel phase between time $0$ and $1$ is exponentially small. Reasoning recursively up to time $\ell$, conditionnally to the fact that at least $\frac{n^2}{2^{i}(1+\epsilon)^i\ell}$ jobs are still alive at time $i$, we conclude that with constant probability a job will survive up to time $\ell\geq \frac{\ln{n^2}}{4\ln\ln{n^2}}$.
\end{proof}

\section{Non-Clairvoyant Batch Set Scheduling}
\label{sec:EE}

We now go back to the general problem. Consider a collection $S=\{S_1,\ldots,S_m\}$ of $m$ sets $S_i=\{J_{i,1},\ldots,J_{i,n_i}\}$ of $n_i$ \PSP\/ jobs, each of them arriving at time zero. The goal is to minimize the setflowtime of the sets.

\subsection{\EE\/ Algorithm}

In the context of the data broadcast with dependencies and for the purpose of proving the competitiveness of their broadcast scheduler, the authors of \cite{RobertSchabanel2007} develop a strategy, namely \EA, for \SP\/ instances (\ie, where each job consists of a sequential phase followed by a fully parallel phase).  The \EA\/ strategy consists in allotting an equal amount $\rho$ of processors to each uncompleted set of jobs and to split arbitrarily (according to some algorithm~$A$) this amount $\rho$ of processors among the uncompleted jobs within each set. This strategy is shown to be $O(1)$-speed $O(1)$-competitive \emph{independently} of the choice of algorithm~$A$, as long as $A$ does not leave some processors unoccupied. It turns out that if the instance is not \SP\/, the choice of $A$ matters to obtain competitiveness. Consider for instance a set of $n$ \PS\/ jobs consisting of a parallel work $\epsilon$ followed by a sequential work $1$ arriving at time~$0$ on one processor; if $A$ schedules the jobs one after the other within the set, the makespan will be $(1+\epsilon)n$ whereas the optimal makespan is $n\epsilon+1$. 

We thus consider the \EE\/ strategy which splits evenly the amount of processors given to each set among the uncompleted jobs within that set. Formally, let $N(t)$ be the number of uncompleted sets at time $t$, and $N_i(t)$ the number of uncompleted jobs in each uncompleted set $S_i$ at time $t$. At time~$t$, \EE\/ on $p$ processors allots to each uncompleted job $J_{ij}$ an amount of processors $\rho_{ij}^t = \frac p{N(t)\cdot N_i(t)}$. Note that in the example above, the makespan of \EE\/ is optimal, $1+n\epsilon$. The following section shows that indeed the competitive ratio of this strategy is asymptotically optimal (up to a constant multiplicative factor). 

\subsection{Competitiveness of \EE}

Scaling by a factor $p$ each sequential work, again we assume w.l.o.g.\/ that $p=1$. 
Consider the \PS\/ instance $S'=\{S'_1,\ldots, S'_m\}$  where $S'_i=\{J'_{i,1},\ldots, J'_{i,n_i}\}$ and each job $J'_{ij}$ consists of a fully parallel phase of work $\Par(J_{ij})$ followed by a sequential phase of work $\Seq(J_{ij})$. Following the proof of Lemma~\ref{lem:S<=S'}, we get:

\begin{lemma} \label{lem:S<=S':SFT}
$\SFT(\EE(S)) \leq \SFT(\EE(S'))$.
\end{lemma}

The next lemmas are the keys to the result. They reduce the analysis of \EE\/ to the analysis of the \emph{flowtime} of \EQ\/ for a collection of \emph{jobs}, which is known from \cite{DBLP:journals/scheduling/EdmondsCBD03} to be $(2+\sqrt3)$-competitive when all the jobs arrive at time~$0$. Let $n = \max_{i=1,\ldots,m} n_i$ be the maximum size of a set $S_i$, and let $\alpha = \frac{(\ln\ln n)^2}{\ln n}$.

\begin{lemma} \label{lem:J=S':SFT}
There exists a \SPS\/ instance $J=\{J_1,\ldots,J_m\}$ of Non-Clairvoyant Batch \emph{Job} Scheduling, such that: $\EQ(J) = \EE(S')[J/S']$, $\Par(J_i)\leq \frac1\alpha \Par(S'_i)$, and ${\Seq(J_i)\leq \frac{\ln n}{\ln(1/\alpha)} \Seq(S'_i)}$, where $\sch[J/S']$ denotes the schedule where $J_i$ receives at any time the total amount of processors alloted to the jobs $J'_{ij}$ of $S'_i$ in schedule $\sch$. 
\end{lemma}

\begin{proof}
Let $\schE = \EE(S')$. Let us construct $J_1$ (the construction of $J_i$, $i\geq 2$, is identical). Consider the jobs $J'_{1,1},\ldots, J'_{1,n_1}$ of $S'_1$ in the schedule $\schE$. Let $t_1=0<\cdots<t_q=c'_1$ (where $c'_1$ denotes the completion time of $S'_1$ in $\schE$), such that during each time interval $[t_k,t_{k+1})$, each job $J'_{1,j}$ remains in the same phase; during $[t_k,t_{k+1})$, the number of jobs of $S'_1$ in a sequential (resp. fully parallel) phase is constant, say $s_k$ (resp. $N_1(t_k)-s_k$).  $J_1$ has $(q-1)$ phases:
\begin{itemize}
\item
if $s_k \geq (1-\alpha)N_1(t_k)$, the $k$-th phase of $J_1$ is sequential of work $w_k = t_{k+1}-t_k$.
\item
if $s_k < (1-\alpha)N_1(t_k)$, the $k$-th phase of $J_1$ is fully parallel of work $w_k = \int_{t_k}^{t_{k+1}} \frac1{N(t)} dt$.
\end{itemize}
$J_1$ is designed to fit exactly in the space alloted to $S'_1$ in $\schE$, thus $\EQ(J) = \schE[J/S']$. We now have to bound the total parallel and total sequential works in $J_1$. Let $K = \{k:s_k\geq(1-\alpha)N_1(t_k)\}$ and $\bar K=\{1,\ldots,q-1\}\smallsetminus K$; by construction, $\Seq(J_1) = \sum_{k\in K} w_k$ and $\Par(J_1) = \sum_{k\in \bar K} w_k$. For each $t\in[t_k,t_{k+1})$ with $k\in \bar K$, the amount of parallel work of jobs in $S'_1$ between $t$ and $t+dt$ is at least $\frac{\alpha N_1(t)}{N(t)\cdot N_1(t)} \,dt = \frac\alpha{N(t)} \,dt$. It follows that  the amount of parallel work of jobs in $S'_1$ scheduled in $\schE$ during $[t_k,t_{k+1})$ is at least $\alpha\int_{t_{k}}^{t_{k+1}} \frac1{N(t)} dt = \alpha \, w_k$. Thus, $\Par(S'_1) \geq \sum_{k\in \bar K} \alpha\, w_k = \alpha \Par(J_1)$, which is the claimed bound. Now, let $A = \cup_{k\in K} [t_k,t_{k+1})$, we have $|A| = \Seq(J_1)$. Since the bound on the size of $A$ in proof of Proposition~\ref{prop:EQ} relies on a counting argument (and is thus independent of the amount of processors given to the set) and the jobs in $S'_1$ are \PS\/, the same argument applies and $|A| \leq \frac{\ln n_1}{\ln(1/\alpha)} \Seq(S'_1) \leq \frac{\ln n}{\ln(1/\alpha)}\Seq(S'_1)$, which conclude the proof.
\end{proof}

\medskip

Let $J'=\{J'_1,\ldots,J'_m\}$ be the \PS\/ instance of Batch Job Scheduling where each job $J'_i$ consists of a fully parallel work of $\Par(J_i)$ followed by a sequential work of $\Seq(J_i)$. Again, as the amount of processors alloted by \EQ\/ to each job is a non-decreasing function of time, pushing parallel work upfront can only make it worse, thus:

\begin{lemma} \label{lem:EQ:J<=J':SFT}
$\EQ(J) \leq \EQ(J')$.
\end{lemma}

We can now conclude on the competitiveness of $\EE$.

\begin{proposition}
\label{prop:EE}
\EE\/ is $\frac{(2+\sqrt3+o(1))\ln n}{\ln\ln n}$-competitive for the setflowtime minimization problem.
\end{proposition}

\begin{proof}
Putting everything together with the analysis of \EQ\/ in \cite{DBLP:journals/scheduling/EdmondsCBD03}:
\begin{align*}
\SFT(\EE(S)) 
	& \leq \SFT(\EE(S')) & \text{(Lemma~\ref{lem:S<=S':SFT})}\\
	& = \FT(\EQ(J)) & \text{(Lemma~\ref{lem:J=S':SFT})}\\
	& \leq \FT(\EQ(J')) & \text{(Lemma~\ref{lem:EQ:J<=J':SFT})}\\
	& \leq (2+\sqrt3) \OPT(J'). & \text{(\cite[Theorem~3.1]{DBLP:journals/scheduling/EdmondsCBD03})}
\end{align*}
Since $J'$ is \PS\/, one can schedule first all the parallel work in $J'$ followed by all the sequential phases together. The flowtime of the resulting schedule is $\Par(J')+\Seq(J')$, thus $\OPT(J')\leq \Par(J')+\Seq(J')$. Finally,
\begin{align*}
\SFT(\EE(S)) 
	& \leq (2+\sqrt3) (\Par(J')+\Seq(J'))\\
	& \leq \textstyle (2+\sqrt3) (\frac1\alpha\Par(S')+\frac{\ln n}{\ln(1/\alpha)}\Seq(S')) & \text{(Lemma~\ref{lem:J=S':SFT})}\\
	& \leq \textstyle(2+\sqrt3) (\frac1\alpha+\frac{\ln n}{\ln(1/\alpha)}) \cdot \max(\Par(S),\Seq(S))\\
	& \leq \textstyle(2+\sqrt3+o(1))\frac{\ln n}{\ln\ln n}\cdot\OPT(S).
\end{align*}
\vspace*{-1.5em}\\
\hspace*{\stretch{1}}~
\end{proof}

\small

\bibliographystyle{plain}
\bibliography{biblio-spaa}

\end{document}